\documentclass{appolb}
\usepackage{graphicx} % Required for inserting images

\usepackage[utf8]{inputenc}
\usepackage{amsfonts,amsmath,amssymb}
\usepackage{bbold}
\usepackage{multicol}
\usepackage{cite}
\usepackage{bigints}
\usepackage{slashed}
\usepackage{tikz}
\usepackage[compat=1.1.0]{tikz-feynman}
\usepackage{bigints}

\usepackage{xcolor}
\usepackage{authblk,ulem}
\usepackage{hyperref} 

\usepackage{graphicx,slashed,bbold,subfigure,amsmath,amssymb,hyperref,enumerate}

\title{An Implicit Regularization Approach to Chiral Models}

\author[1]{Ricardo J. C. Rosado}

\author[2,3]{Adriano Cherchiglia}

\author[4]{Marcos Sampaio}

\author[1]{Brigitte Hiller}

\affil[1]{CFisUC, Department of Physics, University of Coimbra, P-3004-516 Coimbra,
Portugal}
\affil[2]{Instituto de Física Gleb Wataghin, Universidade Estadual de Campinas \\ Rua Sérgio Buarque de Holanda, 777, Campinas, SP, Brasil}
\affil[3]{Departamento de F\'isica Te\'orica y del Cosmos, Universidad de Granada, Campus de Fuentenueva, E–18071 Granada, Spain}
\affil[4]{ Universidade Federal do ABC, 09210-580 , Santo Andr\'e, Brasil}

\begin{document}

\maketitle

%\begin{abstract}
%    We employ implicit regularization (IReg) in quark-antiquark decays of the Z, or of a scalar (CP-even or odd) boson at NLO, and compare with dimensional schemes to reveal subtleties involving infrared divergence cancellation and $\gamma_5$-matrix issues. Besides the  absence of evanescent fields in IReg, such as $\epsilon$-scalars required in certain schemes that operate partially in the physical dimension, we verify that our procedure preserves gauge invariance in the presence of the $\gamma_5$ matrix  without requiring symmetry preserving counterterms while the amplitude is infrared finite as required by the KLN theorem.  
%\end{abstract}

\begin{abstract}
    The decays of the Z boson and CP-even or CP-odd scalar bosons into quark-antiquark pairs have been calculated %The quark-antiquark decays of the Z and of a CP even or odd scalar bosons are obtained
    at NLO in the framework of Implicit Regularization (IReg) , which operates strictly in the physical dimension and complies with the BPHZ procedure. The presence of the $\gamma_5$ matrix is dealt without the need of gauge symmetry restoring counterterms and the Kinoshita-Lee-Nauenberg (KLN) theorem is verified. The results are compared to the ones obtained in the Dimensional Reduction scheme (DRed).
\end{abstract}

\section{Introduction}

In evaluating Feynman amplitudes to address high precision scattering/decay data, regularization and renormalization methods play a major role. Different frameworks have been developed with the intent to ease the increasing complexity encountered in conventional dimensional regularization in higher order processes \cite{gnendiger2017d,torres2021may}. %These encompass new and distinctive (re-) formulations of dimensional schemes, as well as non-dimensional ones \cite{gnendiger2017d,torres2021may}.

In the present contribution Implicit Regularization (IReg) is used to evaluate the aforementioned decays involving chiral vertices at NLO. Surprisingly, although IReg operates strictly in the physical dimension, the need for a suitable treatment of the $\gamma_5$ matrix within divergent integrals stands out. Formally a consistent method for IReg is achieved using a specific dimensional extension \cite{Bruque:2018bmy} in which $\{\gamma_5,\gamma_\mu\}\ne 0$. However in many cases it is possible to operate fully in 4D space, provided adequate steps are undertaken. When a Dirac trace is involved, it often suffices to symmetrise the trace, achieved by using the definition $\gamma_5=i \frac{\epsilon_{\alpha \beta \delta \sigma}}{4!}\gamma^\alpha \gamma^\beta \gamma^\delta \gamma^\sigma$, see e.g. \cite{Viglioni:2016nqc}.%, which bears resemblance to the BMHV scheme, at the price of not fulfilling $\{\gamma_5,\gamma_\mu\}=0$.

The main objective of the present study is to verify under which circumstances the $\gamma_5$ algebra is maintained for an open fermionic line, rendering the intermediate calculational steps as simple as possible, while reproducing the results of more involved schemes. Moreover, in order to show that IReg complies with the finitude theorem of KLN, the fermions are taken to be massless \cite{kinoshita1962mass, lee1964degenerate}.  We refer to \cite{rosado2023infrared} for details.

\subsection{Rules for Implicit Regularisation in a nutshell}

\begin{itemize}
    \item Perform Dirac Algebra
    \item Use the identity 
    \begin{equation}
        \frac{1}{(k \pm p)^2-\mu^2} = \frac{1}{k^2-\mu^2} - \frac{p^2 \pm 2 p\cdot k}{(k^2-\mu^2)[(k \pm p)^2-\mu^2]}
    \end{equation}
    where $\mu^2$ plays the role of an infrared regulator, as many times as necessary to isolate the UV divergent behavior as
    \begin{equation}
        I_{quad}(\mu^2)=\int \frac{1}{k^2-\mu^2} \frac{d^4k}{(2\pi)^2} \text{  
 and   } I_{log}(\mu^2)=\int \frac{1}{(k^2-\mu^2)^2} \frac{d^4k}{(2\pi)^2}
    \end{equation}
    \item A renormalization group scale $\lambda$ is introduced as $\lambda^2\frac{\partial I_{log}(\lambda^2)}{\partial \lambda^2}= -\frac{i}{(4\pi)^2}$. As $\mu^2\rightarrow 0$, $I_{log}(\mu^2)$ parameterizes the IR divergences and $I_{log}(\lambda^2)$ is absorbed by renormalization.
    \item Numerator/denominator consistency and shift invariance is verified in the process of regularization/renormalization \cite{Bruque:2018bmy}.
\end{itemize}

\section{$Z_0$ and (Pseudo)Scalar Calculations}

For the following calculations we'll consider the following variables: $q^\mu$ and $\overline{q}^\mu$ as the momenta of the exiting quark and antiquark respectively; $z^\mu$ and $z=\sqrt{z^2}$ as the momentum of the decaying $Z_0$ boson and its rest mass respectively and for the coupling constants we use $e$ as the fundamental electric charge, $\omega$ the weak mixing angle, $Z_\pm=g_V \pm \gamma_5 g_A$ and $g_V=I_3-Q' sin^2(\omega)$ and $g_A=I_3$ with $I_3$ being the projection of the third component of the isospin of the quarks and $Q'$ their unitary charge. As for the Scalar and Pseudo Scalar we use the Higgs boson and thus the scalar coupling is $\xi_s=\frac{e m_q}{2 sin(\omega) m_W}$ and the pseudocalar coupling $\xi_5=\frac{e I_3 m_q}{sin(\omega) m_W}$ with $m_q$ being the mass of the quarks and $m_W$ the rest mass of the W bosons.

In order to calculate the first order Quantum ChromoDynamics (QCD) corrections to the decay rate of our particles into a quark-antiquark pair we have to consider the following Feynman diagrams in Figure 1.

\begin{figure}
    \centering
    \includegraphics[width=1 \textwidth]{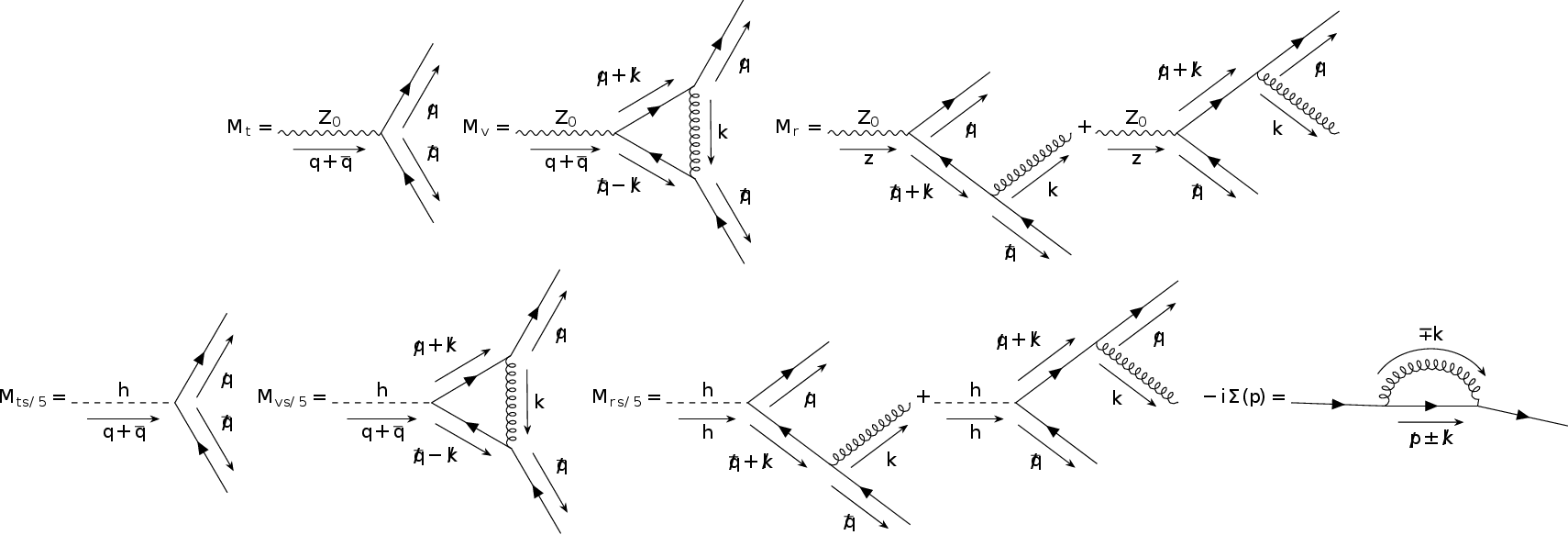}
    \caption{Feynman diagrams corresponding to all contributions to the Amplitude of a $Z_0$ or (pseudo)scalar particle into a quark-antiquark pair up to the Next-to-Leading Order (NLO) in QCD. $M_t, M_{ts/5}$ correspond to tree level, $M_v, M_{vs/5}$ to virtual contributions, $M_r, M_{rs/5}$ to real contributions.}
    \label{fig:enter-label}
\end{figure}

%The decay rate of any particle $\Omega$ of momentum $p^\mu_\Omega$ onto an i number of particles of momenta $q^\mu_i$, is given by given by Fermi's Golden Rule equation:

%\begin{equation}
    %\Gamma = \frac{1}{2m_\Omega} \int \left[\sum_{spin}|M|^2\right] (2\pi)^4\delta(p^\mu-\sum_i q^\mu_i)\prod_i\frac{d^3q_i}{(2\pi)^3 2q_{i0}}
%\end{equation}

\subsection{Tree Level decay rate}

%One first starts with the Tree level, or 0th order, results. The $Z_0$ tree level decay rate ($\Gamma_t$) and the Scalar and Pseudo-Scalar tree level decay rate ($\Gamma_{ts}$ and $\Gamma_{t5}$ respectively) are given by the expressions:

%\begin{equation}
    %M_t = \overline{u}(q) \cdot \frac{-ie \gamma^\mu Z_-}{sin(2\omega)} \cdot v(\overline{q}) \epsilon_\mu (z) \quad \quad \begin{array}{cc}
        %M_{ts}= \overline{u}(q) \cdot -i\xi_s \cdot v(\overline{q}) \\
        %M_{t5}= \overline{u}(q) \cdot -i\xi_5 \gamma_5 \cdot v(\overline{q})
    %\end{array}
%\end{equation}

%the resulting decay rates contributions from these being:

At tree level the well known results \cite{novikov1999theory} are reproduced

\begin{equation}
    \Gamma_t=\frac{e^2(g_V^2+g_A^2)z}{4 \pi sin^2(2\omega)} \quad \quad \quad \quad \Gamma_{ts/5} = \xi_{s/5}^2 \frac{h}{8\pi}
\end{equation}

%showing the well known results, with no anomalous chiral behaviour.

\subsection{Real NLO Contributions}

The amplitudes that are relevant to real contributions to the decay rate of the $Z_0$ and the Higgs (scalar and pseudo scalar decays) are given by the expressions:

%\begin{equation}
    %\begin{array}{cc}
        %\begin{array}{cc}
            %M_{r}=\epsilon_\mu(z) \overline{u}(q) \left[(-ig \gamma^\alpha t^a) \cdot \frac{-i}{\slashed{q}+\slashed{k}} \cdot \frac{-ie \gamma^\mu Z_-}{sin(2\omega)} +\right. \\
            %\left.+ \frac{-ie \gamma^\mu Z_-}{sin(2\omega)} \cdot \frac{i}{\slashed{\overline{q}} + \slashed{k}} \cdot (-ig \gamma^\alpha t^a)\right] v(\overline{q}) \epsilon^*_\alpha (k)
        %\end{array} \\
        %\left\{\begin{array}{cc}
            %M_{rs}=\overline{u}(q) \left(\begin{array}{cc}
                %(-ig \gamma^\alpha t^a) \cdot \frac{-i}{\slashed{q}+\slashed{k}} \cdot -i\xi_s \\
                %-i\xi_s \cdot \frac{i}{\slashed{\overline{q}} + \slashed{k}} \cdot (-ig \gamma^\alpha t^a)
            %\end{array}\right) v(\overline{q}) \epsilon^*_\alpha (k) \\
            %M_{r5}=\overline{u}(q) \left(\begin{array}{cc}
                %(-ig \gamma^\alpha t^a) \cdot \frac{-i}{\slashed{q}+\slashed{k}} \cdot -i\xi_5 \gamma_5 \\
                %-i\xi_5 \gamma_5 \cdot \frac{i}{\slashed{\overline{q}} + \slashed{k}} \cdot (-ig \gamma^\alpha t^a)
            %\end{array}\right) v(\overline{q}) \epsilon^*_\alpha (k)
        %\end{array}\right.
    %\end{array}
%\end{equation}

\begin{equation}
    \begin{array}{cc}
        \begin{array}{cc}
            M_{r}=\epsilon_\mu(z) \overline{u}(q) \left[(-ig \gamma^\alpha t^a) \left(\frac{-i}{\slashed{q}+\slashed{k}}\right) \left(\frac{-ie \gamma^\mu Z_-}{sin(2\omega)}\right) +\right. \quad \quad \quad \quad \quad \quad \\
            \quad \quad \quad \quad \quad \quad \left.+ \left(\frac{-ie \gamma^\mu Z_-}{sin(2\omega)}\right) \left(\frac{i}{\slashed{\overline{q}} + \slashed{k}}\right) (-ig \gamma^\alpha t^a)\right] v(\overline{q}) \epsilon^*_\alpha (k)
        \end{array} \\
        \begin{array}{cc}
            \begin{array}{cc}
                M_{rs}=\overline{u}(q) \left[ (-ig \gamma^\alpha t^a) \left(\frac{-i}{\slashed{q}+\slashed{k}}\right) (-i\xi_s) - \right. \quad \quad \quad \quad \quad \\
                \quad \quad \quad \quad \quad \left. (-i\xi_s) \left(\frac{i}{\slashed{\overline{q}} + \slashed{k}}\right) (-ig \gamma^\alpha t^a) \right] v(\overline{q}) \epsilon^*_\alpha (k)
            \end{array} \\
            \begin{array}{cc}
                M_{r5}=\overline{u}(q) \left[ (-ig \gamma^\alpha t^a) \left(\frac{-i}{\slashed{q}+\slashed{k}}\right) (-i\xi_5 \gamma_5) - \right. \quad \quad \quad \quad \quad \\
                \quad \quad \quad \quad \quad \left. (-i\xi_5 \gamma_5) \left(\frac{i}{\slashed{\overline{q}} + \slashed{k}}\right) (-ig \gamma^\alpha t^a) \right] v(\overline{q}) \epsilon^*_\alpha (k)
            \end{array}
        \end{array}
    \end{array}
\end{equation}

which lead to the decay rate corrections of:

\begin{equation}
    \begin{array}{cc}
        \Gamma_{r}=\Gamma_t \frac{(t^a)^2 g^2}{(4\pi)^2} [2ln^2(\mu_0)-2\pi^2+6ln(\mu_0)+17] \\
        \Gamma_{rs/5}= \Gamma_{ts/5} \frac{(t^a)^2 g^2}{(4\pi)^2} [2ln^2(\mu_0)-2\pi^2+6ln(\mu_0)+19]
    \end{array}
\end{equation}

While there are intermediate IR divergences in this result, parameterized as $ln(\mu_0)$, with $\mu_0=\frac{\mu^2}{m^2_\Omega}$ and with $\Omega$ as the decaying particle, they are expected and will cancel in the total NLO result \ref{tiny} in conformity with the KLN theorem. %and not a product of any anomalous behavior coming from the treatment of the $\gamma_5$.

\subsection{Virtual NLO Contributions}

The amplitudes that are relevant to virtual contributions to the decay rate of the $Z_0$ and the Higgs (scalar and pseudo scalar decays) are given by the expressions:

\begin{equation}
    \begin{array}{cc}
        \begin{array}{cc}
            M_v= \epsilon_\mu(z) \bigintss \overline{u}(q) (-ig\gamma^\alpha t^a) \left(\frac{-i}{\slashed{q}+\slashed{k}}\right) \left(\frac{-ie\gamma^\mu Z_{-}}{sin(2\omega)}\right) \\
            \left(\frac{i}{\slashed{\overline{q}}-\slashed{k}}\right) (-ig\gamma^\beta t^b) \left(\frac{-ig_{\alpha \beta} \delta_{ab}}{k^2}\right) v(\overline{q}) \frac{d^4k}{(2\pi)^4}
        \end{array} \\
        \begin{array}{cc}
            \begin{array}{cc}
                M_{vs}= \bigintss \overline{u}(q) (-ig\gamma^\alpha t^a) \left(\frac{-i}{\slashed{q}+\slashed{k}}\right) (-i\xi_s) \left(\frac{i}{\slashed{\overline{q}}-\slashed{k}}\right) \\
                (-ig\gamma^\beta t^b) \left(\frac{-ig_{\alpha \beta} \delta_{ab}}{k^2}\right) v(\overline{q}) \frac{d^4k}{(2\pi)^4}
            \end{array} \\
            \begin{array}{cc}
                M_{v5}= \bigintss \overline{u}(q) (-ig\gamma^\alpha t^a) \left(\frac{-i}{\slashed{q}+\slashed{k}}\right) (-i\xi_5 \gamma_5) \left(\frac{i}{\slashed{\overline{q}}-\slashed{k}}\right) \\
                (-ig\gamma^\beta t^b) \left(\frac{-ig_{\alpha \beta} \delta_{ab}}{k^2}\right) v(\overline{q}) \frac{d^4k}{(2\pi)^4}
            \end{array}
        \end{array}
    \end{array}
\end{equation}

\noindent leading, after proper regularization, to the corresponding decay rates:

\begin{equation}
    \begin{array}{cc}
        \Gamma_{v}= -\Gamma_t\frac{(t^a)^2 g^2}{(4\pi)^2}[2ln^2(\mu_0)+6ln(\mu_0)+14-2\pi^2] \\
        \Gamma_{vs/5}= -\Gamma_{vs/5} \frac{(t^a)^2 g^2}{(4\pi)^2} [2ln^2(\mu_0)-2\pi^2]
    \end{array}
\end{equation}

While the result for the $Z_0$ here is expected, the results from the Higgs's calculations are not. Recall that the Higgs’s couplings are proportional to the fermion mass, which acquires a contribution at NLO, given by its self-energy.
 %This behaviour has nothing to do with the presence of a $\gamma_5$, as both scalar and pseudo scalar calculations lead to the same result.

\subsubsection{Self-Energy Corrections}

%To correct the apparent anomalous result we remember that the Higgs's couplings contain a mass term.
%By reintroducing the quark mass one has to consider the diagrams that modify the bare mass at the tree level.
%We can use the mass in that term to modify the tree level result to obtain the correct terms. 
%We reintroduce the mass of the quarks which leads to having to consider the diagrams that modify the bare mass.

The correction is given by:

\begin{equation}
    -i\Sigma(\slashed{p})^{(1)}=\int (-ig\gamma^\alpha t^a) \frac{i}{\slashed{p}\pm \slashed{k}-m} (-ig\gamma_\alpha t^a) \frac{-i}{k^2} \frac{d^4k}{(2\pi)^4}
\end{equation}

%By explicitly presenting the mass term in the coupling constant in the tree level equation one can modify it giving us a new term that is first order in QCD. One can can then slot the common mass factor in both terms back into the coupling constant. The resulting complete virtual contributions for the first order virtual decay rate of the higgs (scalar and pseudo scalar) becomes

%This result modifies the implicit mass in the couplings of the tree level creating a new virtual term which when added to the existing ones give.

\noindent which modifies the tree level mass dependence in the coupling, giving rise to a new virtual term. The final expression is

\begin{equation}
    \Gamma_{vs/5}=-\Gamma_{ts/5}\frac{(t^\alpha)^2g^2}{(4\pi)^2}(2ln^2(\mu_0)-2\pi^2+2+6ln(\mu_0))
\end{equation}

\subsection{NLO Decay Rate}

Joining terms and using $\frac{g^2}{4\pi}=\alpha_s$ and $(t^\alpha)^2=C_f=\frac{4}{3}$ one gets

%\begin{equation}
    %\begin{array}{cc}
        %\Gamma_1=\Gamma_t+\Gamma_{1v}+\Gamma_{1r} = \\
        %\Gamma_t \left(1+3\frac{(t^\alpha)^2g^2}{(4\pi)^2}\right) = \\
        %\Gamma_t \left(1+\frac{\alpha_s}{\pi}\right)
    %\end{array} , \begin{array}{cc}
        %\Gamma_{1s/5} = \Gamma_{ts/5} + \Gamma_{ms/5} + \Gamma_{vs/5} + \Gamma_{rs/5} = \\
        %\Gamma_{ts/5} \left(1+17\frac{(t^\alpha)^2g^2}{(4\pi)^2} \right) = \\
        %\Gamma_{ts/5} \left(1+\frac{17 \alpha_s}{3\pi} \right)
    %\end{array}
%\end{equation}

\begin{equation}
    \label{tiny}
    \Gamma_1=\Gamma_t \left(1+\frac{\alpha_s}{\pi}\right) \quad \quad \quad \quad \Gamma_{1s/5}=\Gamma_{ts/5} \left(1+\frac{17 \alpha_s}{3\pi} \right)
\end{equation}

Thus we obtain the well known results \cite{novikov1999theory, braaten1980higgs} for the first order corrected decay rates. % With no anomalous behaviour from the $\gamma_5$ nor any remaining IR divergences.

\section{Dimensional Schemes}

%Dimensional schemes are more prevalent in perturbation calculations of the SM and its extensions. As exaples we have: 't Hooft-Veltman (HV), Conventional Dimensional Regularization (CDR), Dimensional Reduction (DREG) and Four Dimensional Helicity (FDH). The first two are Dimensional Regularization schemes while the later two, called are Dimensional Reduction Schemes, often necessitate the use of evanescent fields in order to be in accordance with unitary \cite{gnendiger2017d}.

%(HV) 't Hooft-Veltman: Internal Vectors are treated in D-dimension, External Vectors are strictly 4-dimensional.
    
%(CDR) Conventional Dimensional Regularization: Both internal and External Vectors are treated in D-dimension.

%(DREG) Dimensional Reduction: Both internal and External Vectors are quasi-4-dimensional.
    
%(FDH) Four Dimensional Helicity: Only The Internal Vectors are Treated in the quasi-4-dimensional space, External Vectors are strictly 4-dimensional.

%Evanescent fields are a Scalar fields that result from the extension of the Vector fields onto the evanescent space \cite{gnendiger2017d}.

%\begin{equation*}
    %d_s=d+n_\epsilon = 4-2\epsilon+n_\epsilon
%\end{equation*}

%\begin{equation}
    %D^\mu_{[d_s]}\psi_i = \partial^\mu_{[d]}\psi_i + i\left(g_s A^\mu_{k[d]} + g_e A^\mu_{k[n_\epsilon]}\right) T^{ijk} \psi_j
%\end{equation}

\subsection{Comparison with FDH (Dimensional Reduction Scheme)}

%One can map the NLO results from the calculations in dimensional schemes onto the IReg scheme. To this one starts by, if there be evanescent fields, doing the transformation $n_\epsilon \rightarrow 2\epsilon$ followed by any numerator/denominator cancellations. One then sets the evanescent couplings to the proper couplings $\alpha_\epsilon \rightarrow \alpha_s$ and converts the divergent terms $\frac{1}{\epsilon} \rightarrow ln(\mu_0)$ and $\frac{2}{\epsilon^2} \rightarrow ln^2(\mu_0)$. Finally one can set all remaining terms of $\epsilon$ to 0.

One can map the NLO results from the calculations in dimensional
schemes onto the IReg scheme, as: $n_\epsilon \rightarrow 2\epsilon$ followed by any numerator/denominator cancellations; $\alpha_\epsilon \rightarrow \alpha_s$, $\frac{1}{\epsilon} \rightarrow ln(\mu_0)$ and $\frac{2}{\epsilon^2} \rightarrow ln^2(\mu_0)$; and finally setting all remaining terms of $\epsilon$ to 0.

\subsection{Scalar Contributions}

The real contributions calculated under FDH are \cite{gnendiger2017d}:

\begin{equation}
    \Gamma^{(v)}_{s/5(FDH)}= \Gamma^{(t)}_{s/5(FDH)} C_f \left[\frac{\alpha_s}{4\pi}\left(-\frac{4}{\epsilon^2}-\frac{6}{\epsilon}-4+2\pi^2\right) + \frac{\alpha_\epsilon}{4\pi}\left(\frac{n_\epsilon}{\epsilon}\right)\right]
\end{equation}

\begin{equation}
    \Gamma^{(r)}_{s/5(FDH)}= \Gamma^{(t)}_{s/5(FDH)} C_f \left[\frac{\alpha_s}{4\pi}\left(\frac{4}{\epsilon^2}+\frac{6}{\epsilon}+21-2\pi^2\right) + \frac{\alpha_\epsilon}{4\pi}\left(-\frac{n_\epsilon}{\epsilon}\right)\right]
\end{equation}

\subsection{$e^-e^+ \rightarrow Z_0 \rightarrow q\overline{q}$ Contributions}

One can also observe the same for the $Z_0$ contribution \cite{gnendiger2017d}:

\begin{equation}
    \sigma^{(v)}_{\gamma(FDH)} = \sigma^{(0)}C_f \left[\frac{\alpha_s}{4\pi}\left(-\frac{4}{\epsilon^2}-\frac{6}{\epsilon}-16+2\pi^2\right) + \frac{\alpha_\epsilon}{4\pi}\left(\frac{n_\epsilon}{\epsilon}\right)\right]
\end{equation}

\begin{equation}
    \sigma^{(r)}_{\gamma(FDH)} = \sigma^{(0)}C_f \left[\frac{\alpha_s}{4\pi}\left(\frac{4}{\epsilon^2}+\frac{6}{\epsilon}+19-2\pi^2\right) + \frac{\alpha_\epsilon}{4\pi}\left(-\frac{n_\epsilon}{\epsilon}\right)\right]
\end{equation}

\section{Conclusions}

\begin{itemize}
    \item It is verified that the KLN theorem is satisfied in our framework.
    \item It is not necessary to introduce evanescent fields, se also \cite{pereira2023higgs,cherchiglia2021two}.%, unlike in partially dimensional methods such as FDH and DRed.
    \item There's a precise matching from dimensional results to IReg at NLO. %We also compared IReg with these methods, showing that, regarding IR divergences, there is a precise matching rule between IReg and dimensional results at NLO.
    \item The $\gamma_5$ right-most-position approach \cite{osti_21503993} is sufficient to render IReg a gauge invariant procedure in this case.% while reproducing the results obtained with more involved schemes literature.
\end{itemize}

\section*{Acknowledgements}

Support from Fundação para a Ciência e Tecnologia (FCT) by projects 10.54499/UIDB/04564/2020 (https://sciproj.ptcris.pt/157582UID), and by 10.54499/UIDP/04564/2020 (https://sciproj.ptcris.pt/157889UID), and grant FCT 2020.07172.BD. A.C. Acknowledges support from National Council for Scientific and Technological Development – CNPq by projects 166523\slash2020-8 and 201013\slash2022-3 and M. S. from CNPq by grant 302790\slash2020-9.

%CERN /FIS-PAR /0040 /2019, CERN /FIS-COM /0035 /2019,
%UID /FIS /04564 /2020,

\addcontentsline{toc}{section}{bibliography}

\bibliographystyle{pnas2009}
\bibliography{bibliography.bib}

\end{document}